\documentclass[fleqn,twoside]{article}
\usepackage{espcrc2}

\usepackage{graphicx}
\usepackage[figuresright]{rotating}

\newcommand{\AmS}{{\protect\the\textfont2
  A\kern-.1667em\lower.5ex\hbox{M}\kern-.125emS}}

\title{New Ideas in Finite Density QCD}

\author{
V. Azcoiti \address[zara]{Departamento de F\'{\i}sica Te\'orica, 
Universidad de Zaragoza, E-50009 Zaragoza, Spain}
        \thanks{Talk presented by V. Azcoiti.},
G. Di Carlo\address[lngs]{INFN, Laboratori Nazionali del Gran Sasso, 
67010 Assergi, (L'Aquila) Italy},
A. Galante\addressmark[lngs]\address[aq]{Dipartimento di Fisica 
dell'Universit\`a di L'Aquila, 67100 L'Aquila, Italy}
V. Laliena\addressmark[zara].
}

\begin{document}

\begin{abstract}
We introduce a new approach to analyze the phase diagram of QCD at finite
chemical potential and temperature, based on the definition of a generalized
QCD action. Several details of the method will be discussed, with particular
emphasis on the advantages respect to the imaginary chemical potential
approach.

\vspace{1pc}
\end{abstract}

\maketitle

In 1997, at the end of the previous century, the general wisdom on the 
possibility of understanding the behavior of matter in extreme conditions 
from first principles was moderately optimistic.
The determination of the grand canonical partition function coefficients of 
QCD by the Glasgow-Illinois group \cite{barbour} using the 
Glasgow reweighting method allowed for the detection of a first order phase 
transition at high baryonic density in the strong coupling regime.

One year later, in 1998, we published a paper containing very surprising 
results. Indeed using 
the microcanonical fermion average approach \cite{mfa} and taking the 
modulus of the 
determinant of the Dirac operator in the integration measure, we were able to 
reproduce with very good agreement \cite{noi1} the 
results obtained with the Glasgow 
method. These results were actually surprising since it was well known that 
the phase of the fermion determinant should play a fundamental role in the 
dynamics of finite density QCD.

In the same year we published another paper \cite{noi2} in which we did a 
detailed 
analysis of the grand canonical partition function and detected how rounding 
effects in the standard routine used by both groups to calculate the 
coefficients of the grand canonical partition function from the eigenvalues 
of the quark propagator matrix, were the responsible for the appearance of 
a phase transition. We proposed also in \cite{noi2} 
a solution to avoid the 
rounding problem and the main conclusion was that the phase transition 
observed by both groups was missing. In other words, the apparent progress 
in the understanding of finite density QCD was fictitious.

From our previous experience we learned that, at large fermion masses, the 
region of the chemical potential in which the phase of the fermion 
determinant becomes almost irrelevant enlarges, and then numerical 
approaches can be successfully 
applied. In this way we were able to predict a tentative phase diagram for 
fat QCD \cite{noi3} which confirmed the theoretical expectations.

This was more or less the general situation of three color QCD until two 
years ago, when an important step forward in the field was possible by using 
two alternative schemes. 
The first scheme, based on extracting information on the phase structure 
of the model at finite $\mu$ from numerical simulations at imaginary chemical 
potential, was implemented in \cite{philippe}. 
The second scheme \cite{fodor} relies on a two-dimensional generalization 
of the Glasgow reweighting method, and has also been combined with 
a Taylor expansion
in the chemical potential which allows simulations at large lattice sizes 
in the small $\mu/T$ region \cite{karsch}.

We want to introduce now a new approach \cite{noi4}
to simulate QCD at finite temperature 
and baryon density. Even if the theory scheme can resemble in some aspects
the imaginary chemical potential approach, its range of applicability seems 
much wider. The results of a test of our 
method using as toy model the Gross-Neveu model at finite $\mu$, as well as 
those for four flavor-QCD, can be found in the talk by Angelo Galante at this 
conference.

Our theory scheme is based on the introduction of a generalized QCD action 
\begin{eqnarray}
S(x,y) = S_{PG} + S_\tau(x, y) + \nonumber \\
{1\over2}\sum^3_{n,i=1}\bar\psi_n \eta_i (n) \left(U_{n,i}
\psi_{n+i} - U^+_{n-i,i}\psi_{n-i}\right)
\label{gaction}
\end{eqnarray}

\noindent
which depends on the two independent parameters $x, y$ trough
\begin{eqnarray}
S_{\tau}(x, y) = \nonumber \\
{1\over2} x \sum_{n}\bar\psi_n \eta_0 (n) \left(U_{n,0}
\psi_{n+0} - U^+_{n-0,0}\psi_{n-0}\right) \nonumber \\
+ {1\over2} y \sum_{n} \bar\psi_n \eta_0 (n) \left(U_{n,0}
\psi_{n+0} + U^+_{n-0,0}\psi_{n-0}\right)
\label{tauaction}
\end{eqnarray}

A natural question now is, what should we expect for 
the phase diagram of this model in the $x, y$ plane?.
The point $x =1$ , $y=0$ would correspond to standard QCD at vanishing 
chemical potential.
Now assume we are in the scaling region but at a physical temperature $T$ 
lower than the deconfining
critical temperature. The point $x = 1$ $y = 0$ will be in
the confined phase. If we increase now the inverse gauge coupling $\beta$,
the physical temperature increases and for $\beta$ values large enough the
point $x= 1$ will be in the unconfined phase. This strongly suggests the
presence of a phase transition point in the $y = 0$ line approaching the
$x = 1$ point in this line by increasing $\beta$ and eventually crossing it
for $\beta$ values large enough.
This argumentation drives us to conjecture the minimal
phase diagram shown in Fig. 1. The solid line would be a line of phase
transitions that crosses the $y= 0$ axis at values of $x$ larger than 1
for small $\beta$ values. By increasing $\beta$ and keeping fixed the
temporal lattice extent $L_t$, the critical point on the
$y= 0$ axis moves toward $x = 1$ and eventually crosses it. The discontinuous
line in this figure stands for the physical line $x^2 - y^2 = 1$ along
which one recovers standard QCD. 
The intersection of the solid line with the discontinuous one will give us
therefore the critical chemical potential of QCD at a given temperature. 

\begin{figure}[t]
\vspace{9pt}
\centerline{\includegraphics*[width=2.3in,angle=270]
{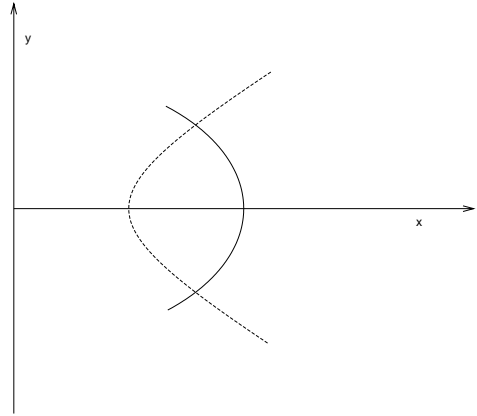}}
\caption{Minimal phase diagram in $(x,y)$ plane.}
\label{fig1}
\end{figure}

The numerical analysis of this model for real values of
$x, y$ is not possible because we find again the sign
problem. However if we take the $y$ parameter as a pure imaginary number,
numerical simulations are feasible. Assuming that the phase transition line 
of  Fig. 1 continues to imaginary 
$y= i\bar y$, the expected phase diagram in the $x, \bar y$ plane will show 
the qualitative structure of Fig. 2, where we have incorporated the 
${2\pi}\over{3L_t}$ periodicity of the model.
We have also included in this figure the line $x^2+\bar y^2 = 1$ 
(dotted line) which
contains the only points accessible to numerical simulations of $QCD$ at
imaginary chemical potential. As stated in the introduction, one can see
now how this approach has in principle more potentialities than the
imaginary chemical potential approach. Indeed by increasing the inverse
gauge coupling $\beta$, the phase transition line of Fig. 2 moves approaching
more and more the origin of coordinates. In some interval $(\beta_m, \beta_M)$
the transition line intersects the dotted line and then a phase transition
will appear at imaginary chemical potential. In such a situation, the physical
temperature is so high that the system is in an unconfined phase for any real
value of the chemical potential. This is in contrast with the fact that, 
within our approach, simulations can be performed at any value of $T$.

\begin{figure}[t]
\vspace{9pt}
\centerline{\includegraphics*[width=2.7in,angle=270]
{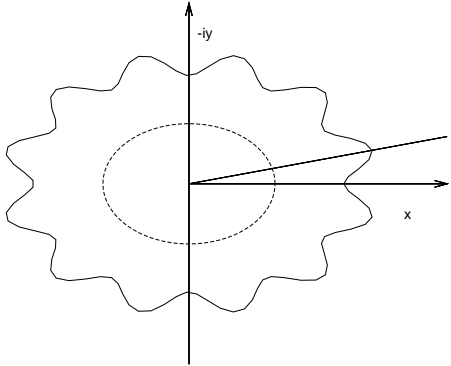}}
\caption{Phase diagram in $(x,\bar y)$ plane.}
\vspace{-9pt}
\label{fig2}
\end{figure}

The practical implementation of our method consists therefore in performing 
numerical simulations at imaginary y to determine the phase transition line 
of Fig. 2, 
and after that, to do analytical extensions to real $y$. This can be done 
mainly in two different ways. 
In the first scheme we fix the inverse gauge coupling and the number of 
temporal lattice points i.e., we fix the physical temperature, and determine 
the equation for the critical line of Fig. 1 as an expansion in 
powers of $y^2$. 
The intersection point of this line with the physical line will give us the 
corresponding critical value of the chemical potential. 
In the second scheme we fix the parameter x and the gauge coupling, and 
look for 
the intersection of the critical line in the $\beta, y$ plane, obtained again 
from analytical extension of the critical line in the $\beta, \bar y$ plane, 
with the 
physical line. This scheme corresponds to fix the value of $\mu/T$, and the 
intersection point will give us the critical temperature at this value of 
$\mu/T$.

The main points to remark in the imaginary chemical potential approach are: 
i. the sign problem is avoided and standard numerical simulations can be 
performed, ii. simulations have to be performed at high temperature 
(large $\beta$) i.e.,
in the deconfined phase for real values of $\mu$, and iii. 
the analytical extension is limited to the range 
${{\mu}\over{T}} < {{\pi}\over{3}}$

Our approach shares the first point with the imaginary chemical potential 
case. Furthermore, and this is a relevant feature, simulations can 
be performed 
at any temperature and any value of $\mu/T$. The price to pay for that is 
one more parameter in the numerical simulations.

A last suggestion, concerning the Glasgow reweighting procedure in two 
parameters space \cite{fodor}, is the possibility of using the fermion 
determinant at the critical point of the $y=0$ line instead of the 
determinant at $\mu=0$ in the integration reweighted 
measure. We suspect that hereby one could improve the overlaps.

\end{document}